\documentclass[aps,prd,twocolumn]{revtex4}
\usepackage{graphicx}

\begin{document}

\title{ Potassium abundance in the Earth and Borexino data}

\author{L.~Bezrukov,$^{1}$ A.~Gromtseva,$^{2}$ I.~Karpikov,$^{1}$ A.~Kurlovich,$^{1}$ A.~Mezhokh,$^{1}$ P.~Naumov,$^{2}$ Ya.~Nikitenko,$^{1}$ S.~Silaeva,$^{1}$ V.~Sinev,$^{1}$ V.~Zavarzina$^{1}$}

\affiliation{$^{1}$ Institute for Nuclear Research of Russian Academy of Sciences, Moscow, Russia}
\affiliation{$^{2}$ National Research Nuclear University MEPHI, Moscow, Russia}

\begin{abstract}
An independent analysis of Borexino single event energy spectrum of recoil electrons and alphas was carried out. We compared two sets of single event sources. The first set is similar to the one used in Borexino Collaboration analysis. The second set additionally includes the scattering of $^{40}$K-geo-($\bar{\nu} + \nu$) on scintillator electrons. 
We found two equivalent minima for $\chi^2$ for second set. The one is for total counting rates $R(^{40}$K-geo-$(\bar{\nu} + \nu)) = 0.0$ and $R(^{210}$Bi) = 10 cpd/100t. The other one is for $R(^{40}$K-geo-$(\bar{\nu} + \nu)) = 7.05$ cpd/100t and $R(^{210}$Bi$) = 6$ cpd/100t. 
We performed MC pseudo-experiments and found that we do not have enough statistics and need to know the bismuth concentration in the scintillator for definite measurement of potassium abundance in the Earth. The possibility of building a next-generation detector for looking for the $^{40}$K-geo-($\bar{\nu} + \nu$) flux is being considered.

\end{abstract}
\maketitle


\section{Introduction}

The Borexino collaboration reported in Ref. \cite{borexsol} on the detection of CNO-cycle solar neutrinos (CNO-$\nu$) and gave the first direct experimental indication of the existence of the CNO cycle in the Sun. This result also showed the way to the solution of the long-standing solar metallicity problem Ref. \cite{borex2}.

The Borexino detector registered CNO-$\nu$ through the neutrino scattering reaction on scintillator electrons. The recoil electron loses its energy and causes a flash of light in the scintillator, the parameters of which are measured. To characterize this flash the scientists use the two equivalent values: the energy loss of charge particles $E$ or the number of photoelectrons recorded by photomultipliers $N_{ph.e.}$.

The Borexino detector achieved uniquely low backgrounds level, which made it possible to distinguish CNO-$\nu$ events.
They used two methods of analysis for the extraction of CNO-$\nu$ events for different energy ranges of recoil electrons: Counting Analysis (CA) and Multivariate Fit (MF).
Borexino formulated the results in terms of total single event counting rate ($R$) initiated by recoil electrons or alphas from certain reaction (for example, CNO-$\nu$ scattering on electrons) without an energy threshold in units cpd/100t $-$ the number of counts per day in 100 tons of scintillator. We will use here the same notation and units.

In CA the Borexino collaboration uses narrow energy range $E$ = 0.74$-$0.85 MeV and obtained the total CNO-$\nu$ scattering rate:
\begin{equation}
R(\rm CNO)_{\rm CA} = (5.6 \pm 1.6)\ {\rm cpd/100t}.
\label{R_CNO_CA}
\end{equation}

In MF they use wide energy range $E$ = 0.32$-$2.64 MeV and obtain the following CNO-$\nu$ scattering rate:
\begin{equation}
R(\rm CNO)_{\rm MF} = (7.2^{+3.0}_{-1.6})\ {\rm cpd/100t}.
\label{R_CNO_MF}
\end{equation}

Expected total CNO-$\nu$ counting rates (Ref. \cite{borex2}) (in Standard Solar Model (SSM) with high (HZ) and low (LZ) metallicity (Ref. \cite{seren1}, Ref. \cite{seren2})), taking into account the MSW-LMA effect (Ref. \cite{holanda}, Ref. \cite{capozzi})) are:
\begin{eqnarray}
R(\rm CNO)_{\rm HZ} = (4.92 \pm 0.78) {\rm cpd/100t},\nonumber \\
R(\rm CNO)_{\rm LZ} = (3.52 \pm 0.52) {\rm cpd/100t}.
\label{R_CNO_EXP}
\end{eqnarray}

The comparison of (\ref{R_CNO_CA}), (\ref{R_CNO_MF}) and (\ref{R_CNO_EXP}) allows to assume the existence of a new additional component contributing to single
Borexino events. It was proposed in Ref. \cite{sinev} to consider the scattering of the $^{40}$K geo-antineutrino and $^{40}$K geo-neutrino ($^{40}$K-geo-($\bar{\nu} + \nu$)) on electrons as such an additional component. This work predicted also the possible counting rate from such process $R(^{40}$K-geo-($\bar{\nu} + \nu)) = (1-4)$ cpd/100t following the Hydridic Earth model or Hydrogen rich Earth model (HE model). The work Ref. \cite{bezruk1} confirmed this result and gave arguments in favor of existence of this rather intensive source.
If potassium abundance is close to the value predicted by Bulk Silicate Earth (BSE) model Ref. \cite{mantov} (0.024\% of the Earth mass) the total potassium counting rate should be $R(^{40}$K-geo-($\bar{\nu} + \nu))\approx$ 0.05 cpd/100t.

The authors of Refs. \cite{bezruk2}, \cite{bezruk4} and \cite{bezruk3} assumed that (\ref{R_CNO_CA}) differs from (\ref{R_CNO_MF}) mainly due to the different contribution of $^{40}$K-geo-($\bar{\nu} + \nu$) events in different energy ranges.

In this article we perform a consistent MF analysis with inclusion of recoil electron energy spectrum from $^{40}$K-geo-($\bar{\nu} + \nu$) scattering.

\section{Energy spectrum of recoil electrons from $^{40}$K-geo-($\bar{\nu} + \nu$)}

$^{40}$K abundance in natural mixture of K isotopes is 0.0117\%. It decays with $T_{1/2} = 1.248\cdot 10^{9}$ y through two modes: to $^{40}$Ca with efficiency 89.28\% emitting $\bar{\nu_{e}}$ with maximal energy 1.311 keV and to $^{40}$Ar emitting one of two possible mono-energetic $\nu_{e}$-s with energies 43.5 keV (10.67\%) and 1.504 keV ($\sim$0.05\%). Decay scheme of $^{40}$K can be found in Ref. \cite{dbase}. Low energy neutrinos cannot be detected but high energy ones can be seen in large volume detector as a small addition to antineutrino's effect.

\begin{figure}[ht]
\begin{center}
\includegraphics[width=86mm]{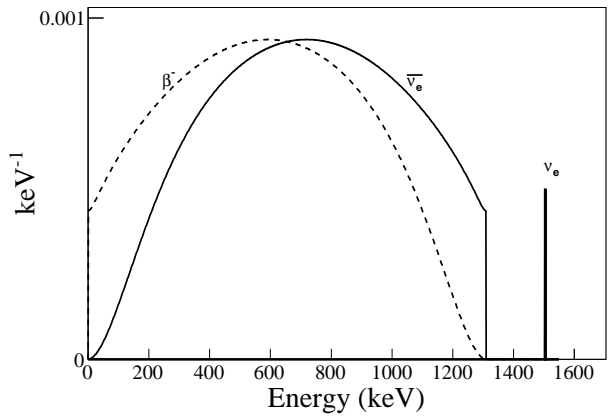}
\end{center}
\caption{\label{figzer} $^{40}$K antineutrinos and neutrinos spectra $-$ solid line. The spectra are normalized to the probablity of emission in decay: 0.8928 for $\bar{\nu_{e}}$-s and 0.0005 for $\nu_{e}$-s. Energy spectrum of electrons emitted in $^{40}$K decay $-$ dashed line.}
\end{figure}

Figure \ref{figzer} demonstrates the total beta, antineutrino and neutrino energy spectra for $^{40}$K decay. The antineutrino spectrum was calculated using the beta one taken in Ref. \cite{dbase}. The beta spectrum is the same as was measured in Ref. \cite{kelly} and calculated in Ref. \cite{mougeot}.

\begin{figure}[ht]
\begin{center}
\includegraphics[width=86mm]{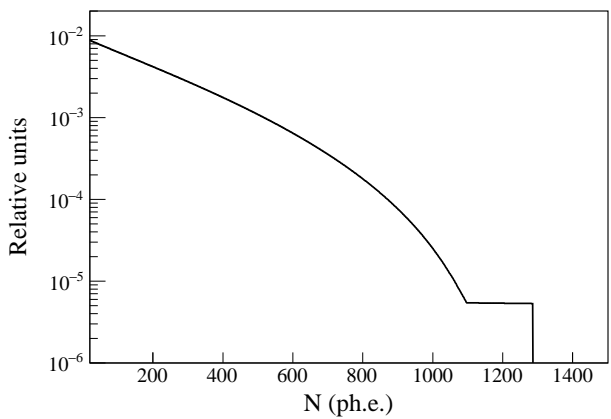}
\end{center}
\caption{\label{figone} The energy spectrum of recoil electrons from $^{40}$K-geo-($\bar{\nu} + \nu$). The pedestal at higher energies corresponds to neutrinos. The total counting rate of recoil electrons for this spectrum is 2.18 cpd/100t.}
\end{figure}

In Figure \ref{figone} one can see the calculated energy spectrum of recoil electrons for $^{40}$K-geo-($\bar{\nu} + \nu$) that follows from the spectrum shape shown in Figure \ref{figzer}. The pedestal at higher energies corresponds to 1.5-MeV neutrinos. The total counting rate of recoil electrons for the spectrum shown in Figure \ref{figone} is equal to 2.18 cpd/100t that corresponds to potassium abundance 1.0\% of the Earth mass.

The energy spectrum normalized to unit is the probability density function and we will use below the notation PDF for such normalized spectrum.

\begin{figure}[ht]
\begin{center}
\includegraphics[width=86mm]{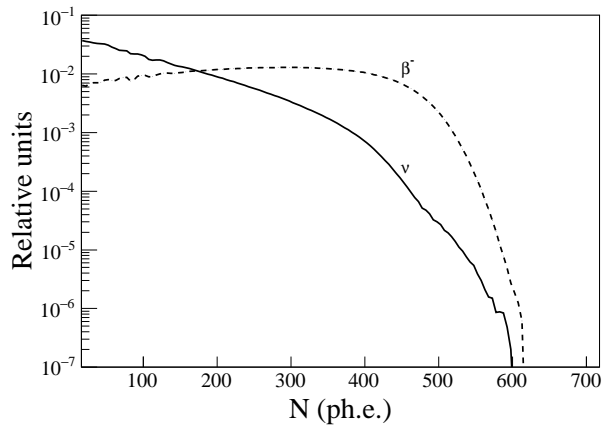}
\end{center}
\caption{\label{figtwo} Recoil electron spectrum from $^{40}$K-geo-($\bar{\nu}$ + $\nu$) transferred to the photoelectron scale according to the algorithm from Ref. \cite{borexphe}$-$ solid line. Energy spectrum of electrons emitted in $^{40}$K decay $-$ dashed line. }
\end{figure}

We will use PDFs in MF as a function of photoelectron number following the Borexino Collaboration because the number of emitted photons in scintillation flash is proportional to the number of recorded photoelectrons.
The calculated recoil electron energy spectrum from $^{40}$K-geo-($\bar{\nu} + \nu$) was transferred from the energy scale to the photoelectron one using the algorithm described in Ref. \cite{borexphe}. In Figure \ref{figtwo} one can find $^{40}$K-geo-($\bar{\nu} + \nu$) PDF used in the MF analysis. $^{40}$K $\beta$ spectrum PDF is also shown here for comparison.

\begin{figure}[ht]
\begin{center}
\includegraphics[width=86mm]{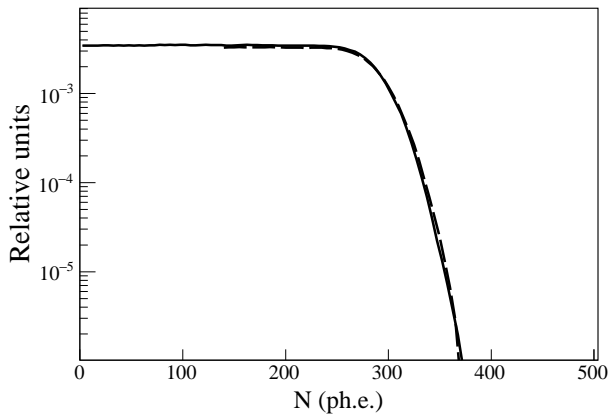}
\end{center}
\caption{\label{figbe} Recoil electron spectrum from $^{7}$Be neutrinos transferred to the photoelectron scale (PDF) according to the algorithm from Ref. \cite{borexphe}. Solid line $-$ our calculation, dashed line $-$ Ref. \cite{borexsol}.}
\end{figure}

To prove that we correctly perform the calculation of $^{40}$K-geo-($\bar{\nu}$ + $\nu$) PDF, one can compare our $^{7}$Be PDF and the one taken from Borexino plot (Ref. \cite{borexsol}) shown in Figure \ref{figbe}.

\section{Energy spectrum of recoil electrons from CNO-$\nu$}

The CNO-$\nu$ spectrum consists of three components: the $^{13}$N, $^{15}$O and $^{17}$F neutrino spectra. The shape of these spectra is well known because the transitions are allowed and can be easily calculated. However, the fluxes depend on the solar model, and many flux predictions exist. In Table \ref{tabl:cnoflux} we can see a number of flux predictions for $^{13}$N, $^{15}$O and $^{17}$F according to different models.

\begin{table}[ht]
\caption{Neutrino fluxes produced in the Sun by nuclei $^{13}$N, $^{15}$O and $^{17}$F in units cm$^{-2}$ s$^{-1}$.}
\label{tabl:cnoflux}
\centering
\vspace{2mm}
\begin{tabular}{ c | c | c | c }
\hline
\hline
Model & $^{13}$N $\times10^{8}$ & $^{15}$O $\times10^{8}$& $^{17}$F $\times10^{6}$ \\
\hline
B16 Ref. \cite{b16} & 5.03 & 1.34 & $<$8.5 \\
GS98 Ref. \cite{gs98} & 3.081 & 2.379 & 5.765 \\
AGSS09 Ref. \cite{agss09} & 2.273 & 1.695 & 3.619 \\
Caffau11 Ref. \cite{caffau11} & 2.801 & 2.123 & 4.648 \\
used in Borexino Ref. \cite{vinoles} & 2.78 & 2.05 & 5.29 \\
\hline
\hline
\end{tabular}	
\end{table}

We have calculated neutrino spectra from $^{13}$N, $^{15}$O and $^{17}$F ourselves in the moment of production inside the Sun. These calculated spectra undergo oscillations according to the MSW-mechanism Ref. \cite{micheev}. As a result each spectrum split in two: the electron neutrino spectrum and $\nu_{\mu}$ + $\nu_{\tau}$ spectrum. For each one we calculated a recoil electron spectrum and composed them in one. 
The sum of three CNO components appear similar to the CNO spectrum used in the Borexino analysis but not exactly. So, we decided to use in our analysis the CNO spectrum digitized from the plot of the Borexino publication. Later we are going to make the analysis using individual CNO spectra. This is needed to check which prediction presented in Table \ref{tabl:cnoflux} better satisfies the experimental data.

\section{Borexino experimental data MF analysis}

The Borexino Collaboration data are open for researchers and available at the site pointed in Ref. \cite{bxopen}. The authors used three-fold coincidence method (TFC), described in Ref. \cite{threef}, to suppress $^{11}$C events; and the data are organized in two files: one is for TFC-subtracted (681.77 d of measurement) events depleted in $^{11}$C and the other one for TFC-tagged (390 d of measurement) events enriched in $^{11}$C. We have obtained these data and used in our analysis the file corresponding to 681.77 days.

We have calculated PDFs for most of the components used in the Borexino analysis and used them in our MF analysis ($^{7}$Be, $pep$, $^{11}$C, $^{210}$Bi, $^{85}$Kr and CNO). Four components were digitized from Ref. \cite{borexsol}, they are: $^{210}$Po alpha-peak, $^{8}$B and two backgrounds caused by external gammas from $^{208}$Tl and the summed spectrum of $^{214}$Bi and $^{40}$K gammas.

The following $\chi^2$ function was used to estimate the goodness of our fit

\begin{equation}
\chi^{2} = \sum_{i=1}^{162} \frac{(N_{i}-\sum_{k=1}^{10}w_{k}\cdot f_{k,i})^2}{\sigma^2_{i}},
\label{chi}
\end{equation}
where $N_{i}$ is the experimental value of counting rate of the Borexino detector in the $i$-th bin of photoelectron number, $w_{k} = R_{k}\cdot 0.713\cdot t_{meas} -$ the weight for the PDF of the $k$-th component with $R_{k} -$ the total counting rate of the $k$-th component in cpd/100t, 0.713 $-$ the ratio of 71.3 t and 100 t and $t_{meas} -$ the measurement time in days, $f_{k,i} -$ the PDF of the $k$-th component and $\sigma_{i}$ is experimental uncertainty.
We used a standard tool for analysis, ROOT 6.22/08, to find the set of parameters $R_{k}\pm\Delta R_{k}$ which minimizes $\chi^{2}$. The $\Delta R_{k}$ corresponds to the confidence level of $68\%$ in ROOT 6.22/08. To check the results of ROOT we used also another minimization program.

The PDF for $^{40}$K-geo-($\bar{\nu}$ + $\nu$) was added to the analysis. So, in total we have 10 PDFs to fit: $^{7}$Be, $pep$, $^{11}$C, $^{210}$Po, $^{210}$Bi, $^{85}$Kr, external backgrounds 1 and 2, CNO and $^{40}$K-geo-($\bar{\nu}$ + $\nu$). Some parameters were constrained as in Ref. \cite{borexsol} $R(pep)=2.74 \pm 0.04$ cpd/100t, $0<R$($^{210}$Bi) $< 11.5\pm 1.3$ cpd/100t, $0<R$($^{85}$Kr) $< 7.5$ cpd/100t, $0 < R$(CNO) $< 10.5$ cpd/100t. We constrained the parameter for $^{40}$K-geo-($\bar{\nu}$ + $\nu$): $0 < R(^{40}$K-geo-($\bar{\nu}$ + $\nu$)) $< 15$ cpd/100t.

At the first stage we made the fit of Borexino data by minimization of $\chi^{2}$ from (\ref{chi}) using our calculated PDFs for the components. We used the energy range from 140 to 940 ph.e. following the Borexino analysis. We will use the notation $\chi^{2}_{glob}$ for $\chi^{2}$ obtained for this energy range. The experimental data were described with a sum of 9 variable components, $^{8}$B was fixed and the parameter for $^{40}$K-geo-($\bar{\nu}$ + $\nu$) was set to zero. As a result we obtained total counting rates that are on the first row of Table \ref{tabl:bor1}. We can observe the similarity of the obtained values of the parameters to the parameters presented by Borexino Collaboration in Refs. \cite{borexsol}, \cite{borexphe} and references inside them.

\begin{figure}[ht]
\begin{center}
\begin{minipage}{86mm}
\includegraphics[width=86mm]{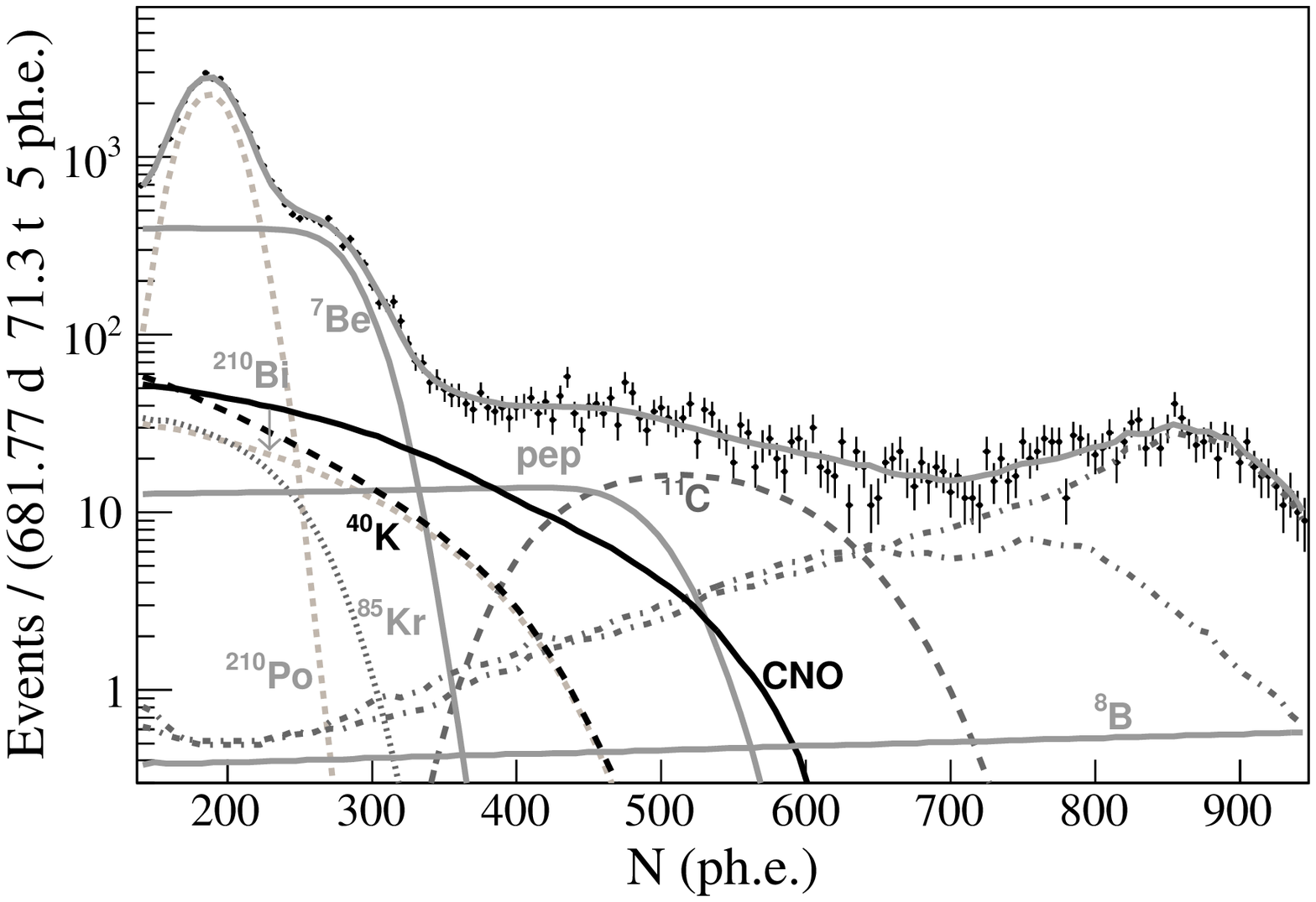}
\end{minipage}\hspace{2pc}%
\begin{minipage}{86mm}
\includegraphics[width=86mm]{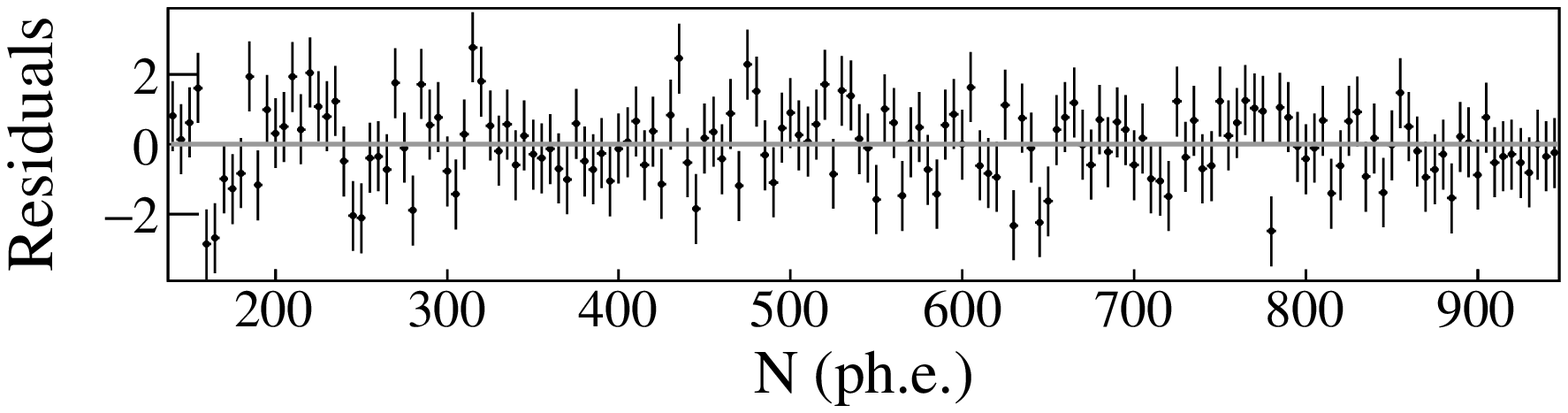}
\end{minipage}
\end{center}
\caption{\label{figthr} Counting rate of recoil electrons and alphas in Borexino detector versus number of photo electrons (points with errors from Ref. \cite{bxopen}). The curves are the contributions of different components of the fit and the sum of all components obtained by the global MF analysis. Black dotted curve is contribution of $^{40}$K-geo-($\bar{\nu} + \nu$) events. The obtained total counting rates for components of the fit are presented in the second row of the Table \ref{tabl:bor1}.}
\end{figure}

The PDF of $^{40}$K-geo-($\bar{\nu} + \nu$) was added at the second stage of the analysis and we made all parameters free. We found two equivalent minima for $\chi^2$. One is for total counting rate $R(^{40}$K-geo-($\bar{\nu} + \nu$)) = 0.0 cpd/100t with the same initial conditions as there were at the first stage and $R_{in.cond.}(^{40}$K-geo-($\bar{\nu} + \nu$)) = 0.0 cpd/100t. The other minimum was for $R(^{40}$K-geo-($\bar{\nu} + \nu$)) = 7.06 cpd/100t with the same initial conditions as the first stage and $R_{in.cond.}(^{40}$K-geo-$(\bar{\nu} + \nu)) > 3.0$ cpd/100t. The result of this fit is shown in Figure \ref{figthr} and obtained total counting rates are in the second row of Table \ref{tabl:bor1}. The fit found reasonable values of total counting rates for all components. We obtained the value of $R(^{40}$K-geo-$(\bar{\nu} + \nu)) = 7.06\pm 7.03$ cpd/100t.
We note here that $\chi^{2}_{glob}$ values obtained at the first and second stages are very close.

We calculated simultaneously the value $\chi^{2}_{loc}$ for the narrow energy range from 240 to 400 ph.e. in which the contribution of $^{40}$K-geo-($\bar{\nu} + \nu$) PDF is the largest.
The obtained values $\chi^{2}_{loc}$ for this narrow energy range are shown in Table \ref{tabl:bor1}. We can see that the difference of $\chi^{2}_{loc}$ between row 1 and row 2 becomes larger than the difference of $\chi^{2}_{glob}$ values in these rows.
However, this difference is not big enough also to make the best choice between the two sets: the one with $^{40}$K-geo-($\bar{\nu} + \nu$) scattering and the one without.

As a result of our MF's we obtained very close values of mean counting rates (and their uncertainties) of the same sources from different sets of the assumed event sources except for $^{210}$Bi and $^{85}$Kr beta-electrons. The inclusion of the scattering of $^{40}$K-geo-($\bar{\nu} + \nu$) on electrons in the analysis mainly reduces the number of events from $^{210}$Bi and $^{85}$Kr.

The PDFs of $^{210}$Bi beta-electrons and recoil electrons from $^{40}$K-geo-($\bar{\nu} + \nu$) and $^{13}$N-$\nu$ (from CNO cycle) are very similar with their end-point energies being practically the same. $E_{max}(^{210}$Bi) = 1160 keV, $E_{max}(^{40}$K-$\bar{\nu}$) $\approx$ 1100 keV, $E_{max}(^{13}$N) = 987 keV. Therefore, they are strongly anti-correlated in MF analysis. However, the PDF of $^{13}$N-$\nu$ is a part of the CNO-$\nu$ PDF which was varied as a whole.

The obtained uncertainty of the total counting rate of $^{40}$K-geo-($\bar{\nu}$ + $\nu$) depends on uncertainties of other parameters. To show this dependence we performed MF with other parameters fixed. The result is in the third row of Table \ref{tabl:bor1}. The uncertainty here is only statistical.

\begin{table*}[ht] 
\caption{Total counting rates obtained without and with the $^{40}$K-geo-($\bar{\nu} + \nu$) component of the fit, cpd/100t.}
\begin{minipage}{\textwidth}
\label{tabl:bor1}
\centering
\vspace{2mm}
\begin{tabular}{ c | c | c | c | c | c | c | c | c | c | c }
\hline
\hline
& $pep$ & $^{7}$Be & $^{11}$C & $^{210}$Po & $^{210}$Bi & $^{85}$Kr & CNO & $^{40}$K & $\chi^{2}_{loc}$ & $\chi^{2}_{glob}$ \\
\hline
1 & 2.78 $\pm$ 0.06 & 46.60 $\pm$ 1.47 & 1.43 $\pm$ 0.13 & 44.83 $\pm$ 0.49 & 9.97 $\pm$ 3.29 & 5.92 $\pm$ 1.98 & 6.20 $\pm$ 2.14 & 0 & 39.282 & 180.921 \\
2 & 2.78 $\pm$ 0.06 & 47.10 $\pm$ 1.51 & 1.41 $\pm$ 0.14 & 44.93 $\pm$ 0.49 & 5.97 $\pm$ 4.00 & 3.76 $\pm$ 2.34 & 6.42 $\pm$ 2.23 & 7.06 $\pm$ 7.03 & 38.546 & 180.830 \\
3 & 2.78 & 47.10 & 1.41 & 44.93 & 5.97 & 3.76 & 6.42 & 7.06 $\pm$ 1.09 & 38.546 & 180.830 \\
\hline
\hline
\end{tabular}
\end{minipage} \hfill
\end{table*}

\begin{table*}[ht]
\caption{Row 1: Parameters and constraints taken for MC simulation of experiment.
Row 2: Result of reconstruction of the real Borexino data with a higher threshold (260 ph.e.) and with parameters and its constraints shown in Row 1 of this Table. Counting rates in cpd/100t.}
\begin{minipage}{\textwidth}
\label{tabl:bor2}
\centering
\vspace{2mm}
\begin{tabular}{ c | c | c | c | c | c | c | c | c | c }
\hline
\hline
& $pep$ & $^{7}$Be & $^{11}$C & $^{210}$Po & $^{210}$Bi & $^{85}$Kr & CNO & $^{40}$K & $\chi^{2}$ \\
\hline
1 &2.7$\div$2.8 & 46$\div$48 &1.3$\div$1.5 &45 &5.5$\div$6.5 &3.5$\div$4.5 &$>$0 &$>$0 &\\
2 & 2.78$\pm$0.05 & 47.8$\pm$1.6 & 1.41$\pm$0.14 & 45$\pm$0.5 & 5.5$\pm$0.7 & 3.5$\pm$0.7 & 6.65$\pm$1.6 & 6.97 $\pm$ 4.6 & 131.4 \\
\hline
\hline
\end{tabular}
\end{minipage} \hfill
\end{table*}

\begin{table*}[ht]
\caption{Result of the MF using the $^{40}$K $\beta$-spectrum instead of the antineutrino one, cpd/100t.}
\begin{minipage}{\textwidth}
\label{tabl:bor3}
\centering
\vspace{2mm}
\begin{tabular}{ c | c | c | c | c | c | c | c | c | c | c }
\hline
\hline
& $pep$ & $^{7}$Be & $^{11}$C & $^{210}$Po & $^{210}$Bi & $^{85}$Kr & CNO & $^{40}$K & $\chi^{2}_{loc}$ & $\chi^{2}_{glob}$ \\
\hline
1 & 2.78 $\pm$ 0.05 & 46.61 $\pm$ 1.45 & 1.42 $\pm$ 0.13 & 44.83 $\pm$ 0.49 & 9.91 $\pm$ 2.85 & 5.93 $\pm$ 2.16 & 6.23 $\pm$ 1.89 & 0.0 $\pm$ 0.24 & 39.312 & 180.921 \\
\hline
\hline
\end{tabular}
\end{minipage} \hfill
\end{table*}

\section{Simulation of the experiment}

We performed MC simulated experiments to understand the influence of the finite sample of data, the parameter constraints and the other conditions of the fit on reconstructed results. 
The MC simulated experiments help also to understand the possibilities of a next generation Borexino-type detector to observe the $^{40}$K-geo-$(\bar{\nu} + \nu)$ flux.

We took a set of PDFs with certain fixed counting rates and simulated the Borexino-like spectrum many times via the MC method. Each time we had a new finite sample of Borexino-like data. We reconstructed the values of counting rates using the procedure (\ref{chi}) described in Section IV for each sample and, as a result, obtained probability distributions of the reconstructed counting rates. We obtained wide non-symmetric distributions with their mean values different from the values fixed in MC mainly for components with low counting rates($R(^{210}$Bi), $R(^{85}$Kr), $R(^{40}$K)). The main reason for this is the effect of statistical fluctuations of components with high counting rates. These fluctuations affect the reconstructed counting rates of components with low counting rates. The value of this effect depends on statistics. Therefore, we found that the reconstructed counting rates in the Borexino-like data samples obtained using this procedure can have systematic biases relative to their true values.

Below we will give an example of our pseudo-experiments.
We have two equivalent sets of components after our MFs of the experimental Borexino data sample. One set is in the first row of Table \ref{tabl:bor1} and another set is in the second row of Table \ref{tabl:bor1}. The main difference is the reconstructed $R(^{40}$K-geo-($\bar{\nu} + \nu)$) and $R(^{210}$Bi) rates.

We chose for our example of pseudo-experiments the parameters from second row of Table \ref{tabl:bor1}. Knowing about the existence of systematic biases, we took for MC simulation of data samples the counting rates $R(^{40}$K-geo-($\bar{\nu} + \nu)$) = 5 cpd/100t and $R$(CNO) = 5 cpd/100t instead of the values in the second row of Table \ref{tabl:bor1}. We simulated $10^4$ pseudo-experiments.
  
For all simulated spectra we reconstructed the parameters using of the procedure (\ref{chi}) described in Section IV. We decided for these pseudo-experiments to exclude polonium events by applying this procedure to smaller energy range with a higher threshold of 260 ph.e. (260 - 940).
Only the following two parameters were free: $R(^{40}$K-geo-($\bar{\nu} + \nu)$) and $R$(CNO). The other parameters were fixed or constrained, see the first row of Table \ref{tabl:bor2}.

\begin{figure}[ht]
\begin{minipage}{86mm}
\includegraphics[width=86mm]{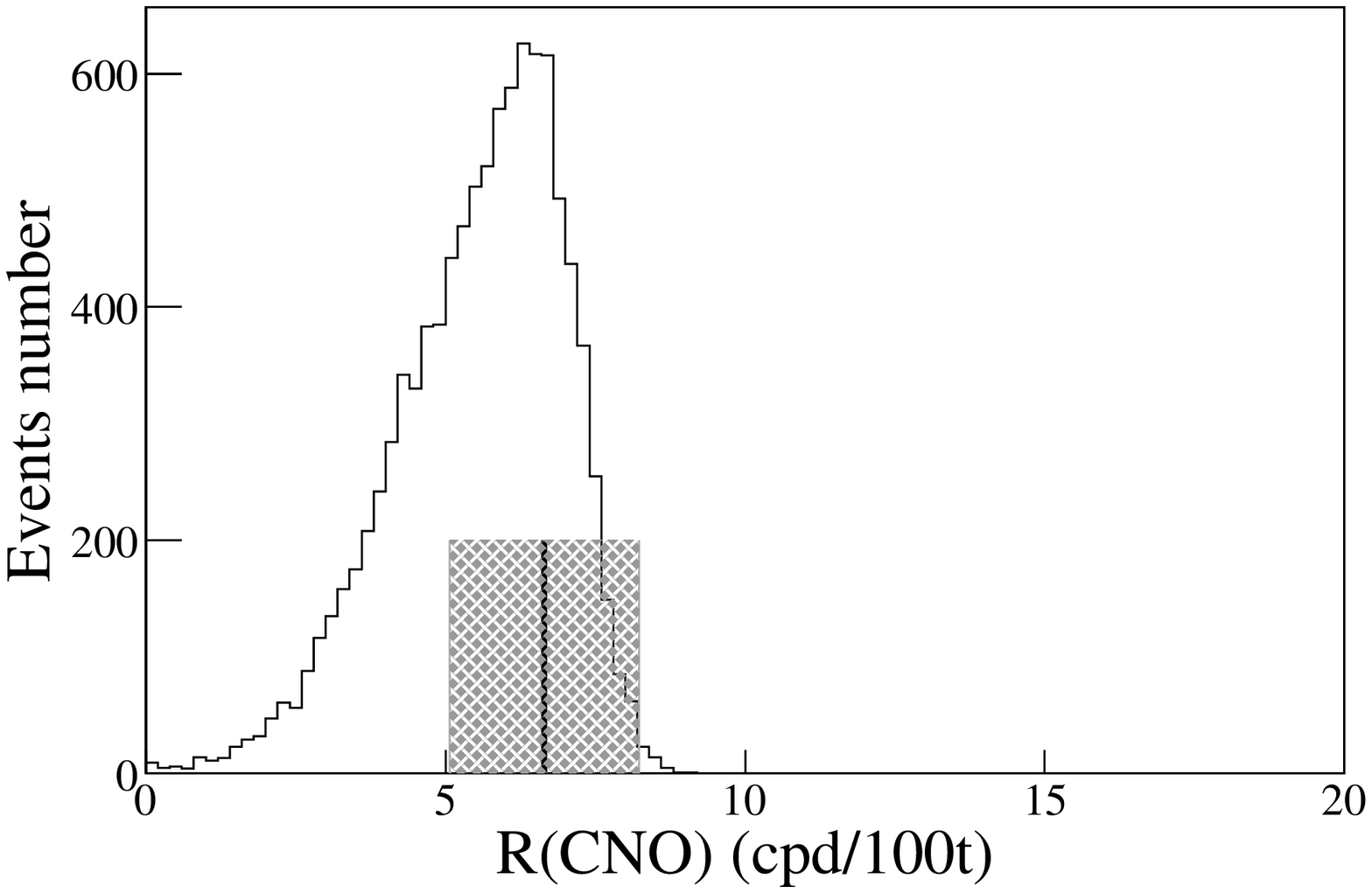}
\end{minipage}\hspace{2pc}%
\begin{minipage}{86mm}
\includegraphics[width=86mm]{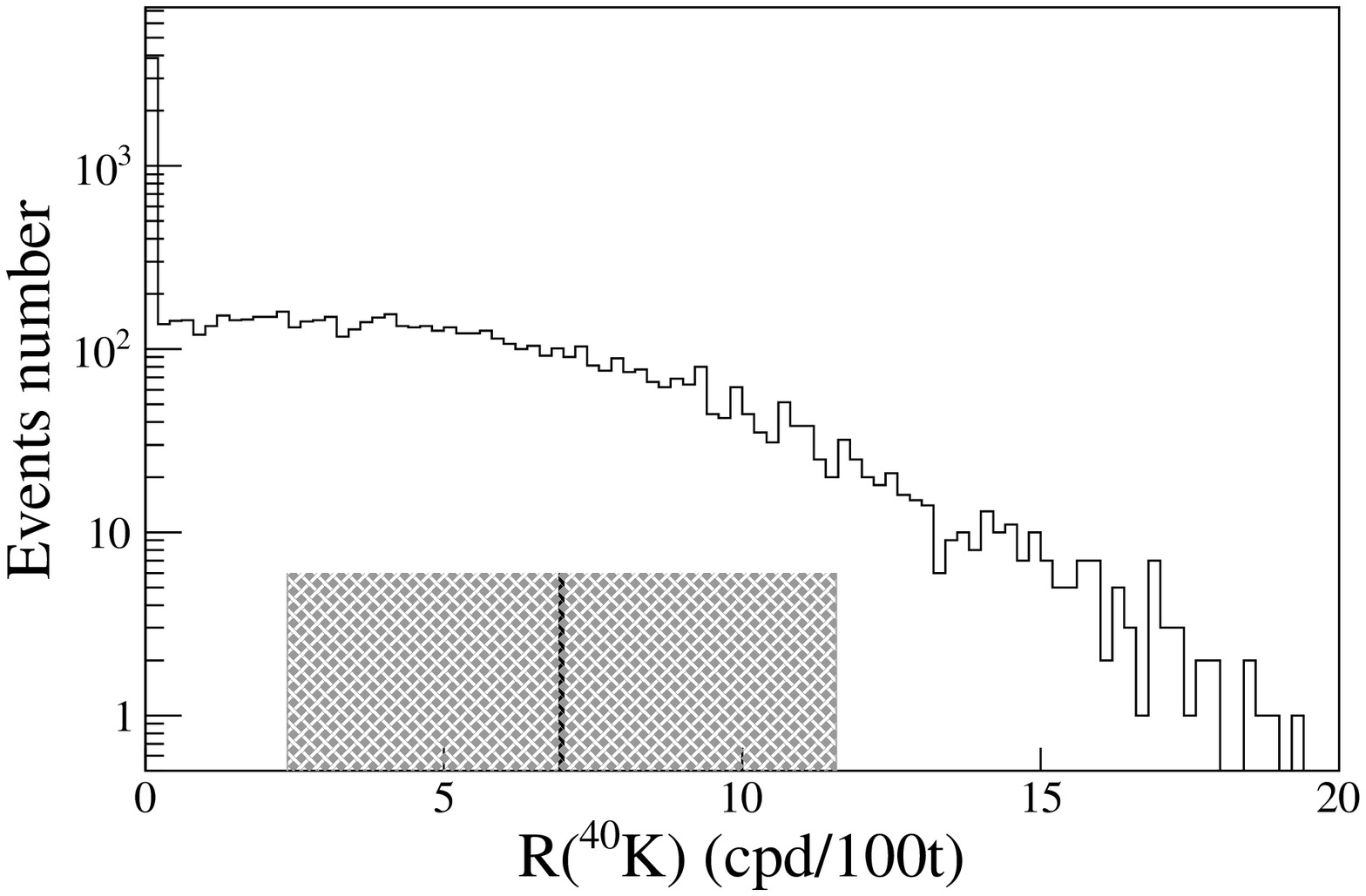}
\end{minipage}
\caption{\label{figforMC} Reconstructed counting rate distributions for $10^4$ pseudo-experiments. The up panel plot shows the $R$(CNO) distribution, the down panel one is the $R(^{40}$K-geo) distribution. The shaded areas correspond to reconstructed values (mean $\pm 1\sigma$) $R(^{40}$K-geo-$(\bar{\nu} + \nu))$ and $R$(CNO-$\nu$) using the experimental Borexino spectrum and the procedure (\ref{chi}) applied to the energy range of 260 to 940 ph.e.}
\end{figure}

Figure \ref{figforMC} shows the reconstructed counting rate distributions for $10^4$ pseudo-experiments. The up panel contains the $R$(CNO) distribution, the down one has the $R(^{40}$K-geo) distribution.

Note that the $R(^{40}$K-geo) distribution has  $3 267$ pseudo-experiments with $R(^{40}$K-geo) = 0.0. 

We obtained mean values $\bar{R}$(CNO)$ = 5.55 \pm1.40$ cpd/100t and $\bar{R}(^{40}$K-geo)$ = 3.58 \pm3.77$ cpd/100t instead of $R$(CNO) = 5 cpd/100t and $R(^{40}$K-geo) = 5 cpd/100t taken for MC simulations. This result demonstrates that the reconstructed mean values are biased.

Shaded areas in Figure \ref{figforMC} correspond to reconstructed values $R(^{40}$K-geo-$(\bar{\nu} + \nu))$ and $R$(CNO-$\nu$) from the experimental Borexino spectrum via the procedure (\ref{chi}) with parameters and constraints shown in Row 1 of Table \ref{tabl:bor2} and applied to the smaller energy range of 260 to 940 ph.e. The obtained values of parameters showed in the Row 2 of Table \ref{tabl:bor2} are similar to the corresponding values in Row 2 of Table \ref{tabl:bor1}.

We will use here the upper limit on the total $^{40}$K-geo-$(\bar{\nu} + \nu)$ counting rate to characterize the distribution on Figure \ref{figforMC}. 
The upper limit is equal to $R(^{40}$K-geo-$(\bar{\nu} + \nu)) < 8.6$ cpd/100t for 90\% CL.
 This value corresponds to the Earth's potassium abundance of 3.9\% by weight.

The upper limit value depends on the $^{210}$Bi content in the scintillator. The total counting rate $R(^{210}$Bi$) = 5.5\div6.5$ cpd/100t for distribution \ref{figforMC}.  If there is less bismuth, then the limit will increase. If there is more bismuth, then the limit will decrease. For example, if the bismuth counting rate $R(^{210}$Bi$) = 10$ cpd/100t, then the limit will be close to zero. 

Recall here that the set of parameters that we used to obtain Figure \ref{figforMC} is the most likely. 
Nevertheless, this analysis shows that we cannot reject the HE model and overhand the current data set recorded by the Borexino detector does not have enough statistics to measure definitely the value of $^{40}$K-geo-($\bar{\nu} + \nu)$ flux.

\section{Next generation Borexino-type detector}

Our analysis showed that the potential capabilities of the Borexino-type detector are not exhausted. 

Let consider a new detector of the Borexino-type with lower backgrounds and higher statistics. 
This require the low background nylon for detector inner vessel, better energy resolution, and detector must be deeper placed. 

The pure nylon inner vessel will result in a larger fiducial volume and smaller amount of $^{210}$Po emanating into the scintillator. The latter will make it possible to measure the concentration of $^{210}$Bi in the scintillator by studying the Low Polonium Field in a stationary scintillator.

Knowing the exact value of the bismuth content in the scintillator will reduce the errors of the reconstructed values of the $^{40}$K-geo-($\bar{\nu} + \nu)$ counting rate. Compare the $^{40}$K value errors in row 2 of Table \ref{tabl:bor1} and in row 2 of Table \ref{tabl:bor2}.

The result of the MC simulated experiments with the constrained $^{210}$Bi content in the scintillator $R(^{210}$Bi$) = 5.5\div6.5$ cpd/100t is given in Figure \ref{figforMC} for finite samples (681.77 d $\times$ 71.3 t). As statistics increase, all distributions become narrower and biases decrease. It can be seen that statistics should be larger for a definite observation of $^{40}$K-geo-($\bar{\nu} + \nu)$ flux. 

Finally, we should like to demonstrate  how the procedure (\ref{chi}) can react on inclusion in the set of event sources an incorrect PDF instead of $^{40}$K-geo-($\bar{\nu} + \nu$) PDF.
We have changed $^{40}$K-geo-($\bar{\nu} + \nu$) PDF to the energy spectrum of electrons emitted in $^{40}$K decay  (dashed line on Figure \ref{figtwo}) and performed the same fit. In case of incorrect PDF usage the $R(^{40}$K-$\beta)$ parameter becomes zero and $\chi^{2}_{loc}$ slightly increases. This result is shown in Table \ref{tabl:bor3}.

\section{Conclusion}

We carried out our own analysis of the Borexino experimental data published in Ref. \cite{borexsol} and available through Ref. \cite{bxopen}.

The following conclusions can be drawn from our MF and MC analyses:

\begin{itemize}
\item We calculated the recoil electron energy spectrum from $^{40}$K-geo-($\bar{\nu} + \nu$) scattering on electrons and transferred it to the observed photoelectron scale.

\item 
We performed the MF's of the experimental sample (681.77 days, energy range from 140 to 940 ph.e.) of the Borexino single event energy spectrum for two sets of event sources: with and without $^{40}$K-geo-($\bar{\nu} + \nu$) scattering on electrons.

\item 
As a result of our MF's we obtained mean counting rates (and their uncertainties) very close to the corresponding Borexino values for the same sources from different event source sets except for the ones for $^{210}$Bi and $^{85}$Kr. The inclusion of $^{40}$K-geo-($\bar{\nu} + \nu$) scattering on electrons in the analysis mainly reduces the obtained $^{210}$Bi and $^{85}$Kr counting rates.

\item We found two equivalent minima for $\chi^2$ for the case with inclusion of $^{40}$K-geo-($\bar{\nu} + \nu$) events . The one is for total counting rates $R(^{40}$K-geo-$(\bar{\nu} + \nu)) = 0.0$ and $R(^{210}$Bi) = 10 cpd/100t. The other one is for $R(^{40}$K-geo-$(\bar{\nu} + \nu)) = 7.05$ cpd/100t and $R(^{210}$Bi$) = 6$ cpd/100t. 

\item Our MF analysis of the experimental sample of Borexino single events showed that we cannot make best choice between the sets with and without $^{40}$K-geo-($\bar{\nu} + \nu$) scattering.
\item The analysis of MC finite samples of Borexino-like single events allowed us to establish an upper limit for the total potassium counting rate $R(^{40}$K-geo-$(\bar{\nu} + \nu)) < 8.6$ cpd/100t for 90$\%$ CL under condition that the total bismuth counting rate is $R(^{210}$Bi$) = 5.5\div6.5$ cpd/100t. This value corresponds to the Earth's potassium abundance 3.9\% by weight. This figures show that the Borexino data does not allow us to reject the Hydridic Earth model.

\item Our analysis showed that the potential capabilities of the Borexino-type detector are not exhausted. It is necessary to build a new detector of the same type as the Borexino, but having lower backgrounds and higher statistics. This means that, it should have the inner vessel made of low background nylon, better energy resolution, and be deeper placed.

\end{itemize}

\vspace{2mm}
\section*{Acknowledgments}

We are grateful to G. V. Sinev for valuable discussions on transferring energy loss to observable number of photoelectrons and idea to generate pseudo-experiments of the Borexino data, to F. L. Bezrukov for discussions and fruitful remarks and to I. I. Tkachev for common support and the possibility to discuss the results in his seminar.


\begin{thebibliography}{25}


\bibitem{borexsol} M.~Agostini {\it et al}. (Borexino Collaboration), Experimental evidence of neutrinos produced in the CNO fusion cycle in the Sun, Nature, {\bf 587}, 577 (2020); arXive: 2006.15115 [hep-ex].

\bibitem{borex2} M.~Agostini {\it et al}. (Borexino Collaboration), Sensitivity to neutrinos from the solar CNO cycle in Borexino, arXive: 2005.12829 [hep-ex].

\bibitem{seren1}
N.~Vinyoles, A.~M.~Serenelli, F.~L.~Villante, S.~Basu, J.~Bergstr\"om, M.~C.~Gonzalez-Garcia, M.~Maltoni, C.~Pe\~na-Garay, and N.~Song, A new Generation of Standard Solar models, Astrophys. J. {\bf 835}, 202 (2017); arXiv:1611.09867.

\bibitem{seren2} F.~L.~Villante and A.~M.~Serenelli, An update discussion of the solar abundance problem, arXiv: 2004.06365.

\bibitem{holanda} P.~C.~De Holanda, W.~Liao, and A.~Yu.~Smirnov, Toward precision measurements in solar neutrinos, Nucl. Phys. B {\bf 702}, Nos. 1–2, 307 (2004).

\bibitem{capozzi}
F.~Capozzi, E.~Di~Valentino, E.~Lisi, A.~Marrone, A.~Melchiorri, and A.~Palazzo, Unfinished fabric of the three neutrino paradigm, Phys. Rev. D {\bf 104}, 083031 (2021).

\bibitem{sinev} V.~V.~Sinev, L.~B.~Bezrukov, E.~A.~Litvinovich,
I.~N.~Machulin, M.~D.~Skorokhvatov, and S.~V.~Sukhotin, Looking for antineutrino flux from $^{40}$K with large liquid scintillator detector, Phys. Part. Nucl. {\bf 46}, 186 (2015).

\bibitem{bezruk1}
L.~B.~Bezrukov, I.~S.~Karpikov, A.~S.~Kurlovich, A.~K.~Mezhokh, S.~V.~Silaeva, V.~V.~Sinev, and V.~P.~Zavarzina, On the contribution of the 40K geo-antineutrino to single Borexino events, arXiv: 2004.02533.

\bibitem{bezruk2} L.~B.~Bezrukov, I.~S.~Karpikov, A.~S.~Kurlovich, A.~K.~Mezhokh, S.~V.~Silaeva, V.~V.~Sinev, and V.~P.~Zavarzina, On first detection of solar neutrinos from CNO cycle with Borexino, arXiv: 2007.07371v2.

\bibitem{bezruk4} L.~B.~Bezrukov, V.~P.~Zavarzina, I.~S.~Karpikov, A.~S.~Kurlovich, A.~K.~Mezhokh, S.~V.~Silaeva, and V.~V.~Sinev, Interpretation of First Detection of Solar Neutrinos from CNO Cycle with Borexino, Bull. Russ. Acad. Sci. Phys. {\bf 85}, 566 (2021).

\bibitem{bezruk3} V.~Sinev, L.~Bezrukov, I.~Karpikov, A.~Kurlovich, A.~Mezhokh, S.~Silaeva and V.~Zavarzina, What can the CNO neutrinos flux measurement done by Borexino say about $^{40}$K geoneutrino flux?, J. Phys.: Conf. Ser. {\bf 1690}, 012170 (2020).

\bibitem{dbase} www-nds.iaea.org/relnsd/vcharthtml/VChartHTML.html

\bibitem{kelly} K.~A.~Kelley, G.~ B.~Beard, and R.~A.~Peters, Nucl. Phys. {\bf 11}, 492 (1959).

\bibitem{mougeot} X.~Mougeot, BetaShape: A new code for improved analytical calculations of beta spectra, EPJ Web Conf. {\bf 146}, 12015 (2017).
\bibitem{mantov} W.~F.~McDonough, {\it The Mantle and Core}, edited by C. R. W. (Elsevier-Pergamon, Oxford), Vol. 2 of Treatise on Geochemistry, 547 (2003); G.~Fiorentini, M.~Lissia and F.~Mantovani, Geoneutrinos and Earth's interior, Phys. Rept. {\bf 453}, 117 (2007).

\bibitem{borexphe} M.~Agostini {\it et al}. (Borexino Collaboration), Simultaneous precision spectroscopy of pp, $^7$Be, and pep solar
neutrinos with Borexino Phase-II, Phys. Rev., D {\bf 100}, 082004 (2019).

\bibitem{b16} C.~Fröhlich and J.~Lean, The Sun's Total Irradiance: Cycles and Trends in the Past Two Decades and Associated Climate Change Uncertainties, Geophys. Res. Lett Geophys. Res. Lett., {\bf 25}, 4377 (1998).
\bibitem{gs98} N.~Grevesse and A.~J.~Sauval, Standard solar composition, Space Sci. Rev., {\bf 85}, 161 (1998).

\bibitem{agss09} M.~Asplund, N.~Grevesse, A.~J.~Sauval and P.~Scott, The Chemical Composition of the Sun, Ann. Rev. Astron. Astrophys., {\bf 47}, 481 (2009).

\bibitem{caffau11} E.~Caffau, H.-G.~Ludwig, M.~Steffen, B.~Freytag and P.~Bonifacio, Solar Chemical Abundances Determined with a CO5BOLD 3D Model Atmosphere, Sol. Phys., {\bf 268}, 255 (2011).

\bibitem{vinoles} L.~Ludhova, talk at TAUP-2021, https://indico.ific.uv.es/event/6178/contributions/15981 /attachments/9341/12521/Ludhova-SolarsGeonu\_TAUP2021\_final.pdf; N.~Vinoles {\it et al}, Astrophys. J., {\bf 836}, 202 (2017).

\bibitem{micheev} S.~P.~Mikheyev, A.~Yu.~Smirnov, Resonance Enhancement of Oscillations in Matter and Solar Neutrino Spectroscopy, Sov. J. Nuc. Phys., {\bf 42} (6), 913 (1985); Yad. Fiz. {\bf 42}, 1441 (1985); L.~Wolfenstein, Phys. Rev. D {\bf 17} (9), 2369 (1978).

\bibitem{bxopen} https://borex.lngs.infn.it

\bibitem{threef} G.~Bellini et al. (Borexino Collaboration), First evidence of pep solar neutrinos by direct detection in Borexino, Phys. Rev. Lett. {\bf 108}, 051302 (2012).

\bibitem{Greenwood} N.~N.~Greenwood, A.~Earnshaw, {\it Chemistry of the Elements}, 2nd ed. (Butterworth-Heinemann), 1359 (1997).

\end{thebibliography}
\end{document}